\documentclass[aps,prd]{revtex4}

\usepackage{latexsym}
\usepackage{epsfig}
\usepackage{scrextend}
 \usepackage[french]{babel}

\usepackage{url}
\usepackage{hyperref}
\usepackage{latexsym}
\usepackage{epsfig}
\usepackage{scrextend}
\usepackage{lipsum}

\usepackage{natbib}

\usepackage{tikz}
\usetikzlibrary{positioning,arrows,decorations.pathmorphing,arrows.meta}

\usepackage{xcolor}
\usepackage{amssymb}
\usepackage{amsmath}
\usepackage{graphicx}

 \definecolor{blue}{RGB}{7,80,201}  
 \definecolor{red}{RGB}{200,20,1}

\def\be{\begin{equation}}
\def\ee{\end{equation}}
\def\bea{\begin{eqnarray}}
\def\eea{\end{eqnarray}}
\def\ba{\begin{array}} 
\def\ea{\end{array}}
\def\bc{\begin{center}}
\def\ec{\end{center}}

\def\ghost#1{}

\def\simge{\mathrel{%
   \rlap{\raise 0.511ex \hbox{$>$}}{\lower 0.511ex \hbox{$\sim$}}}}
\def\simle{\mathrel{
   \rlap{\raise 0.511ex \hbox{$<$}}{\lower 0.511ex \hbox{$\sim$}}}}

\def\dis{\displaystyle}

\newcommand{\footremember}[2]{%
    \footnote{#2}
    \newcounter{#1}
    \setcounter{#1}{\value{footnote}}%
}

\begin{document}

\title{\boldmath \Large MICROSCOPE\, limits \vspace{2mm}  for new long-range forces
\vspace{5mm}
\hbox{and implications for unified theories}}

\vspace{3mm}

\author{{\sc P}{ierre} {\sc Fayet}
\vspace{3mm} \\ \small }

\affiliation{Laboratoire de physique th\'eorique de l'\'Ecole normale sup\'erieure\vspace{.3mm}\\ 24 rue Lhomond, 75231 Paris cedex 05, France 
\hspace{-1mm}\footremember{a}{\vspace*{0mm}   \em CNRS UMR 8549, \vspace{0mm}ENS, PSL University \,\em \& \em UPMC, \vspace{.8mm}Sorbonne Universit\'es}
\vspace{1mm}\\
\hbox{and \,Centre de physique th\'eorique, \'Ecole polytechnique, 91128 Palai\-seau cedex, France}
\vspace{2mm}\
\vspace{1mm}}


\begin{abstract}

\textwidth 10cm
\textheight 26cm
{
\vspace{2mm}
\hspace*{-.5mm} Many theories beyond the Standard Model involve an extra $U(1)$ gauge group.
The resulting gauge boson $U$, in general mixed with the $Z$ and the photon, may be massless or very light, and very  weakly coupled.
It may be viewed as a generalized dark photon interacting  with matter through a linear combination  $[\,\epsilon_Q Q + \epsilon_B B+\epsilon_L L\,]\,e\,$, \,involving $B\!-\!L\,$ in a grand-unified theory,
\linebreak pre\-su\-mably through $B\!-\!L-.61\,Q$, inducing effectively a very small repulsive force between neutrons.
\vspace{-2mm}\\
\hspace*{2.5mm} This new force, if long-ranged, may manifest through apparent violations of the Equivalence Principle. They are approximately proportional  to $\epsilon_B+\epsilon_L/2$, times a combination involving mostly $\epsilon_L$.  New forces coupled to $B-L$ or $L$ should lead to nearly opposite values of the E\"otv\"os parameter $\delta$, and to almost the same limits for $\epsilon_{B-L}$ or $\epsilon_{L}$, as long as no indication for $\delta\neq 0$ is found.
\vspace{1.5mm\-}\\
\hspace*{2.5mm} 
We derive new limits from the first results of the MICRO\-SCOPE experiment testing the Equivalence Principle in space.
A long-range force coupled to $[\,\epsilon_Q Q + \epsilon_{B-L} (B-L)\,]\,e\,$
or $[\,\epsilon_Q Q + \epsilon_L L\,]\,e\,$
should verify $\,|\,\epsilon_{B-L}|$ or  $|\,\epsilon_L|  < .8\ 10^{-24}$, and a force coupled to $[\,\epsilon_Q Q +\- \epsilon_B B\,]\,e\,$, $|\epsilon_B| <  5\ 10^{-24}$.
\vspace{1.5mm}\\
\hspace*{2.5mm} 
We also discuss, within supersymmetric theories, how such extremely small gauge couplings $g"$, typically $\simle 10^{-24}$, may be related to a correspondingly large $\,\xi"D"$ term associated with a huge initial \linebreak vacuum 
energy density, $\propto \!1/g"^2$. The corresponding hierarchy between energy scales, 
by a factor 
$\propto\! 1/\sqrt {g"}\simge 10^{12}$,  involves a very large scale  $\sim 10^{16}$ GeV, that  may be associated with inflation, or super\-symmetry breaking with a very heavy gravitino, leading to possible values of $\delta$ within the expe\-ri\-mentally accessible range.
\ghost{
\\
We also discuss how, in supersymmetric theories, such extremely small gauge couplings $g"$, typically $\simle 10^{-24}$, may be related to a huge initial vacuum energy density involving a large hierarchy between energy scales, by a factor 
\linebreak
$\simge 10^{12}$, involves a very large  scale $\sim 10^{16}$~GeV, as associated with inflation, or supersym\-me\-try- breaking with a very heavy gravitino, leading to possible values of $\delta$ within the experimentally accessible  range.
}
 \vspace{1mm} \\
\hspace*{2mm}
\vspace{-2.5mm}\\
\hbox{\hspace{110mm}LPTENS/17/33}
\vspace{-2.5mm}\\
}
\end{abstract}

\maketitle
 
\textheight 26cm

\section{\boldmath A new long-range force from an extra $U(1)$} 
\label{sec:1}

\,

\vspace{-.5mm}

Are all fundamental particles and forces now known? It would be presumptuous to think so.
In particular, extremely weak new long-range forces could exist, adding their effects to those of gravity. We derive here, from the first results  of the MICROSCOPE experiment  testing the Equivalence Principle in space  \cite{micro}  \footnote{MICRO-Satellite \`a tra\^in\' ee Compens\' ee pour l'Observation du Principe d'Equivalence.}, improved limits on the intensity of such new forces, as compared to gravity, or electromagnetism. We also discuss how such forces could fit within fundamental theories involving grand-unification, supersymmetry, and inflation.

\vspace{2mm}

While the Standard Model has been  confirmed brilliantly with the discovery of the Brout-Englert-Higgs boson, 
many physics questions remain unanswered, dealing in particular with 
dark matter and dark energy, the hierarchy of mass scales and interaction strengths, the problems of quantum gravity and the quest for a possible unification of all interactions.
Gravity is very weak as compared to the other interactions, the gravitational attraction between two protons being smaller than their electrostatic repulsion by 
\vspace{-2.5mm}
a factor
\be
\frac{\hbox{gravity}}{\hbox{electromagnetism}}\ =\ \frac{G_N m_p^2}{e^2/{4\pi\epsilon_\circ}}\,=\, \frac{1}{\alpha}\ \left(\frac{m_p}{m_{\hbox{\scriptsize Planck}}}\right)^{\!2}\ \simeq \ .8093\ 10^{-36}\,.
\,
\ee
Why is gravity so weak, or equivalently why is the Planck energy, 
$
E_{\hbox{\scriptsize Planck}}\,=\, \sqrt{{\hbar c^5}/{G_N}}\,\simeq\, 1.2209\ 10^{19}$ $ \hbox{GeV},
$
so large,
still remains a mystery.
Another aspect of the question is that gravity should become strong at very high energies, with an effective coupling 
$\,\propto G_NE^2/\hbar c^5= (E/E_{\hbox{\footnotesize Planck}})^2$,
generally making a quantum theory of gravity inconsistent. Approaches towards a consistent theory of quantum gravity, and most notably  string theories, usually involve additional $U(1)$  symmetries and describe many new fields and particles, some of them extremely weakly coupled. This may also occur in supersymmetric theories, possibly in connection with grand-unification, supersymmetry breaking, supergravity and the  vacuum energy density, which plays an important role in the evolution of the Universe.

\vspace{2mm}
Irrespectively of their possible origin, the couplings of a spin-1 particle, hereafter denoted as the $U$ boson,  are generally expected to obey a gauge symmetry principle. The $U$ mass may vanish if this gauge symmetry is conserved, or be naturally small, especially if the corresponding gauge coupling $g"$ is very small or extremely small. 
If the new symmetry $U(1)_U$ associated with the $U$ is spontaneously broken, e.g.\,\,through the v.e.v.\,\,of an extra singlet field,   the $U$ acquires a mass $m_U$, vanishing with $g"$, 
mediating a new force 
\vspace{-4mm}
of range
\be
\label{range}
\lambda_U\,=\,\frac{\hbar}{m_U  c}\,\simeq \,1\,973\ \hbox{km}\ \left( \frac{m_U}{10^{-13}\ \hbox{eV}/c^2}\right)^{-1}.
\ee

We have long discussed such extensions of the standard model with a gauge group  enlarged to 
\be
G= SU(3)\times SU(2)\times U(1)_Y\times \hbox{extra-} U(1)_F\,,
\ee
and a small or very small extra-$U(1)$ gauge coupling $g"$
\cite{plb80,npb81,plb86,plb89,newint}.
Assuming that the three families of quarks and charged leptons acquire their masses from a single electroweak doublet $\varphi$ 
as in the standard model, or several but with the same gauge quantum numbers (as for $h_1^c$ and $h_2$ in the Supersymmetric Standard Model), the gauge invariance of their Yukawa couplings requires the new $U(1)_F$ quantum number $F$ 
to be 
\vspace{-2mm}
expressed as \cite{plb89,newint}
\be
\label{f}
\ba{ccc}
\ \ \ \ \ \ \ \ \ F&=&\!\alpha_B \,B\,+\,\beta_i \,L_i\,+\,\gamma \,Y\ \,[+\,F_d]\,.
\ea
\ee
The $U(1)_F$ current is expressed for quarks and leptons as a combination of  the weak-hyper\-charge $Y$ current  with the $B$ and $L$ currents. 
This current may also include a possible dark-matter \cite{npb04} or extra spin-0 singlet contribution associated with a ``hidden sector'', providing the extra term $F_d$ in (\ref{f}). We also ignore, at this moment, a possible contribution from the $R$-symmetry current in a supersymmetric theory, as we are dealing mainly with the usual spin-$\frac{1}{2}$ quarks and leptons, with $R=0$, disregarding super\-partners.
\,The $Y$ term, if present in (\ref{f}) (for $\gamma\neq 0$), is responsible for a mixing of the extra-$U(1)$ gauge field with the neutral electroweak ones, ultimately leading to a contribution in the $U$ current proportional to the electromagnetic current.

\vspace{2mm}
 In the framework of grand-unification \cite{gg,gqw} the simultaneous appearance of $B,\,L$ and $Y$ in 
(\ref{f}) is necessary to ensure that the new $U(1)_F$ gauge group commutes with the $SU(5)$ grand-unification group.
The $U(1)_F$ quantum number (normalized to $\gamma=1$, for $\gamma\neq 0$)  may then be expressed as \cite{newint,plb89}
\be
\label{f2}
F= Y- \frac{5}{2}\ (B-L)\ \ [+\,F_d]\ .
\ee
This $F$ is equal to $- 1/2$ and $+ 3/2$ for the $SU(5)$ {\,\boldmath $\underline {10}$'s\, and $\,\underline {\bar 5}$}'s \,of quarks and leptons, and  $+1$ for the BEH quin\-tu\-plet {\boldmath $\underline 5$} \,describing the electroweak doublet $\varphi$ responsible for their masses, respectively. The quantity 
\vspace{-4mm}
expressed as 
\be
B-L\,=\,  \frac{2}{5} \ \,(Y-F)\ ,
\ee
in the visible sector,  remains conserved at this stage, as long as no neutrino Majorana mass terms are considered, with the grand-unification gauge bosons $X^{\pm 4/3}$ and $Y^{\pm 1/3}$ having 
$\,B\!-\!L= \frac{2}{5} \, Y= \pm \,\frac{2}{3}\,$.
 
\vspace{2mm}
Independently of this possibility of grand-unification, the v.e.v. of the spin-0 doublet $\varphi$, taken with $F=Y=1$ (for $\gamma\neq 0$),  induces the electroweak breaking in a way involving also the extra-$U(1)$ gauge field $C$, 
according to
\be
\underbrace{SU(3)\times SU(2)\times U(1)_Y}_{\hbox{\small possibly  within \normalsize $\,SU(5)$}}\  \times\   \hbox{extra-} U(1)_F\,\to\ SU(3) \times U(1)_{QED}\times U(1)_U\,.
\ee
 The three neutral gauge fields  $\,W_3,\,B$ and $C$ get mixed into the massless photon field $A$, the massive $Z$ field and a new neutral field $U$, still massless at this stage. They are given by \cite{plb89,epjc}
\vspace{.5mm}
\be
\label{33}
\framebox [12.6cm]{\rule[-1.7cm]{0cm}{3.6cm} $ \dis
\left\{\, \ba{ccccc}
A\!&=&\!\hbox{$\dis \frac{g' \,W_3+g\,B}{\sqrt{g^2+g'^2}}$}\!&=&\,\sin\theta \, W_3+\cos\theta \, B\,,
\vspace{2mm}\\
Z\!&=&\!  \hbox{$\dis \frac{g \,W_3-g'B-g''C}{\sqrt{g^2+g'^2+g''^2}}$}\!&\simeq&\,\cos\theta \, W_3-\sin\theta \, B\,, 
\vspace{2mm}\\
U= Z \times A \!&=&\, \hbox{$\dis \frac{g''(g \,W_3-g'B)+(g^2+g'^2)\,C}{\sqrt{g^2+g'^2}\,\sqrt{g^2+g'^2+g''^2}}$}\,&\simeq&\,C\ ,
\ea
\right.\,
$}
\ee
with $\,m_W=gv/2\,$ and $\,m_Z=\sqrt{g^2\!+\!g'^2\!+\!g"^2}\ v/2\simeq m_W/\cos\theta\,$.
 The $U$ field is obtained
  from a mixing, with a very small angle $\xi$, 
  \vspace{-3mm}
  given by  \,\footnote{The sign of $g"$ may always be defined as positive (or if one wishes so, negative), thanks to the possibility of redefining the extra-$U(1)$ gauge field $C$ through a change of sign, $C\to -\,C$.}
  \be
  \tan\xi=\frac{g"}{\sqrt{g^2+g'^2}}\ ,
  \ee
 between the extra-$U(1)$ gauge field $C$ and  the usual weak neutral gauge field $\,Z_\circ= \cos\theta \, W_3-\sin\theta \, B\ $.
  \vspace{2mm}

The $U$ is coupled to a current  ${\cal J}_U^\mu$, 
  \vspace{-.4mm}
combination of the extra-$U(1)$ current $J^\mu_F$ with  the standard weak neutral current $J_3^\mu-\sin^2 \theta \,J_{\rm em}^\mu$,
which remains conserved, as long as the $U$ stays massless. 
Its axial part cancels out in this limit, as required for a conserved current. It is expressed as a combination of the baryonic, leptonic (or $B-L$) and {electromagnetic} currents, according to \cite{plb89,epjc}
\be
\label{curr2}
\ba{ccl}
{\cal J}_U^\mu\!\!&=&\!\hbox{\small$\dis \frac{g"}{2}$} \cos\xi\,\left(\,J^\mu_Y+ \alpha_B J^\mu_{B}+\beta_i\,J^\mu_{L_i} \ [ +J_d^\mu ]\,\right)+ \sqrt{g^2+g'^2}\, \sin\xi\  (J_3^\mu-\sin^2 \theta J_{\rm em}^\mu)
\vspace{2mm}\\
&=&\! 
e\tan\chi \,\left(J^\mu_{\rm em}+
\hbox{\small$\dis \frac{1}{2\,\cos^2\theta}$}\ (\alpha_B \,J^\mu_{B}+\beta_i\,J^\mu_{L_i} \ [+ J_d^\mu]\,) \right),
\ea
\ee
with the coupling 
\be
\label{chi}
\epsilon_Q \,e\,= \,e\,\tan\chi= \,g"\cos^2\theta\,\cos\xi \,\simeq \,g"\cos^2\theta\,.
\ee

\vspace{1mm}

This includes and generalizes, within an extended electroweak or grand-unified theory,  the  very specific case of a ``dark photon'' coupled 
proportionally to  electric charges. It also includes the case of a new massless gauge boson coupled to baryon number \cite{ly}.
 The $U$ boson may be viewed, more generally, as a generalized dark photon  coupled  to Standard Model particles as in (\ref{curr2}),
through the linear combination
\be
\label{qu}
(\,\epsilon_Q Q + \epsilon_B B+\epsilon_L L\,)\,e\,,
\ee
 \vspace{-6mm} 
 
\noindent 
with
\,\footnote{In the special case $\gamma=0$ for which $Y$ does not contribute to $F$, the $U$ does not mix with the elec\-tro\-weak gauge bosons, so that $\xi=0$ with $\epsilon_Q=0$, still allowing for  $\epsilon_B e= (g"/2)\,\alpha_B$ and $\epsilon_{L_i} e=(g"/2)\,\beta_i$\,.}
\be
\label{b-l}
\epsilon_Q \,e\,=\,g"\cos^2\theta\,\cos\xi \,,
\ \ \
\epsilon_B \,e\,= \,\hbox{\small$\dis \frac{g"}{2}$}\,\alpha_B\,\cos\xi\,,\ \ \ \epsilon_{L_i}e\,= \,\hbox{\small$\dis \frac{g"}{2}$}\,\beta_i\,\cos\xi\,\ \ \
(\hbox{where}\ \cos\xi\,\simeq\,1)\,.
\ee
The $U$ may stay massless or acquire a mass (e.g. from the v.e.v. of an extra singlet field), possibly extremely small, making the new force finite-ranged without affecting significantly its couplings (only by extremely small terms  $\propto m_U^2/m_Z^2$).

\vspace{2mm}

The corresponding new force acts {\it additively} on ordinary neutral matter, proportionally to a linear combination of baryon and lepton numbers. 
This is in practice equivalent to considering a force acting effectively on a linear combination of the numbers of protons and neutrons, $Z=L$ and $N=B-L\,$. Such a new force is thus generally expected to be {\it repulsive} (except if $\epsilon_B B+\epsilon_L L$  has different signs for the Earth and the test mass considered).
This is in contrast with a spin-0 induced force, normally expected to be attractive, and not expected to have such an additivity property, making its couplings more difficult to evaluate \cite{damour}. This may allow to distinguish the spin-1 and spin-0 induced cases, should such a force be found.

\section{The new force within grand-unification}

Furthermore, within grand unification the v.e.v.\,\,of the spin-0 quintuplet $\varphi$ breaks the $SU(5)\times U(1)_F$ symmetry into a $SU(4)_{\rm es}\times U(1)_U$ subgroup, with $SU(4)_{\rm es}\supset \,SU(3)_{\rm QCD}\times U(1)_{\rm QED}$ appearing as an  {\it electrostrong} \,symmetry group unifying directly electromagnetic with strong interactions and commuting with $U(1)_U$ at the grand-unification scale (at which $\cos^2\theta=5/8$) \cite{epjc}. \,As for $SU(5)$ itself, this $SU(4)$ electrostrong symmetry group is spontaneously broken to $SU(3)_{\rm QCD}\times U(1)_{\rm  QED}$ by the v.e.v. of an adjoint spin-0 field, or possibly through the compactification of  an extra dimension, then leading to
a grand-unification scale of the order of the compac\-ti\-fi\-cation scale, $m_X\approx \pi\hbar/\hspace{.3mm}Lc$ \cite{f87}.

\vspace{2mm}

The extra $U(1)_F$ quantum number $F$ is given by (\ref{f2}), so that $B$ and $L$ contribute to the $U$ current in (\ref{curr2}) only through their difference $B-L$\,.
This appearance of $B-L$ also follows from the requirement of anomaly-cancellation, in the presence of $\nu_R$ fields and ignoring a possible contribution from $L_i-L_j$.   
The coupling (\ref{qu}) then reads in a grand-unified theory,
\be
\label{gut}
\epsilon_{B-L}\,\left[ \,(B-L)-\hbox{\small$\dis \frac{4}{5}$}\,\cos^2\theta\,Q\,\right]\,e\,,
\ee
i.e. 
\vspace{-3mm}
approximately
\be
\label{gut2}
\framebox [4.8cm]{\rule[-.25cm]{0cm}{.7cm} $ \dis
\epsilon_{B-L}\ (B-L - .61 \ Q)\ e\,
$}
\ee
using $\sin^2\theta\simeq .238$ at low energy.
\,$\epsilon_{B-L}$ and $\epsilon_Q$ are related 
\vspace{-.5mm}
by
\be
\label{gbl}
\epsilon_{B-L}\,e\,=\, -\,\hbox{\small$\dis \frac{5}{4\,\cos^2\theta}$} \ \epsilon_Q\,e\,= \,-\,\hbox{\small$\dis \frac{5}{4}$}\ g"\cos\xi\,
\simeq\, -\,\hbox{\small$\dis \frac{5}{4}$}\ g"\,\simeq\,-\,1.64\ \epsilon_Q\,e\simeq\,-\,.497\ \epsilon_Q\,,
\ee
as also seen from (\ref{b-l}). The $Y$ and $B-L$ terms in expression (\ref{f2}) of $F$ are responsible, after the electroweak breaking, for the couplings of the $U$ boson both to the electric charge (as for a pure dark photon) and to $B-L\,$, respectively.
The expression  $\,\epsilon_{B-L}\,e=\,-\,(5/4)\,g"\cos\xi \,$  fixing the coupling of the $U$ to $B-L$ originates directly from the extra-$U(1)_F$ gauge coupling $g"/2$
 (from $iD_\mu= i\partial_\mu- ... - \frac{g"}{2} F\,C_\mu$),  \,times the coefficient $\,-\,5/2\,$ of $\,B-L\,$ within the $U(1)_F$ quantum number $F$ in (\ref{f2}), times  $\,\cos\xi\simeq 1$ from the $C.U$ scalar product in the $3\times 3$ mixing matrix (\ref{33}).

\vspace{2mm}

At the grand-unification scale, for which $\cos^2\theta =5/8$, expression (\ref{gut}) of the coupling would read \cite{epjc}
\be
\label{estrong}
\hspace{21mm} \epsilon_{B-L}\,\left[\, (B-L)-\hbox{\small$\dis \frac{Q}{2}$}\ \right]\,e\, \ \ \ \ \ 
\hbox{ (invariant under the \normalsize$SU(4)_{\rm es}$ {\it electrostrong} {\small \,symmetry)}}
\ee
evolving into $\,\epsilon_{B-L}\,(B-L - .61 \,Q)\,e\,$ at low energy.
The $SU(4)_{\rm es}$-symmetric expression (\ref{estrong}) vanishes
 for the $u, \, c$ and $t$ quarks, as necessary since $\,u_L$ and $\,\bar u_L$, $\,c_L$ and $\,\bar c_L$, $\,t_L$ and $\,\bar t_L$,  join into  $SU(4)_{\rm es}$ sextets 
 of the electro\-strong symmetry group unifying directly photons with gluons. (For the same reason the $u,\,c$ and $t$ couplings to the $Z$ are purely axial for $\sin^2\theta=3/8$,
 \vspace{-.3mm}
  as required by the electrostrong symmetry.)
 \linebreak
 At the same time also $B-L-\frac{Q}{2}\,$ has the same value $-1/2\,$ for the $\bar d$ and $e$ fields, 
  \vspace{-.3mm}
  joining into $(\bar d,e)$, $(\bar s, \mu)$, and $(\bar b,\tau)$ vectorial Dirac $SU(4)_{\rm es}$ antiquartets. 
\,$\epsilon_{B-L}$, equal to $\,-\,2\,\epsilon_Q\,$ at the grand-unification scale, may thus be already present at this scale
\footnote{This mixing leading to a non-vanishing $\epsilon$ occurs already at the GUT scale despite the absence of a non-diagonal term $\propto C_{\mu\nu} B^{\mu\nu}$, \,that would ``kinetically mix'' the abelian $U(1)_F$  field $C^\mu$ with the weak-hypercharge $B^\mu$, non-abelian in the grand-unification framework. $B^\mu$ and $C^\mu$ always appear as orthogonal fields, still allowing for a non-vanishing $\epsilon_Q=\tan\chi$ as seen from eqs.\,(\ref{curr2},\ref{chi}).},
without having to be generated by radiative corrections.

\vspace{2mm}
The above coupling (\ref{gut},\ref{gut2}) simplifies for neutral particles into \cite{plb89,newint,epjc}
\be
\epsilon_{B-L}\,(B-L)\,e\,,
\ee
sufficient for a phenomenological analysis. As the new force then acts in opposite ways on protons and electrons
it involves in practice the number of neutrons, $B-L= A-Z=N$, effectively acting as a repulsive force between neutrons. The resulting apparent violations of the Equivalence Principle will then be proportional to the difference between the ratios $N/A_r$ for the elements constituting the two test masses ($A_r$ being the relative atomic mass of the element considered, scaled to 12 for a $^{12}$C atom), leading to an E\" otv\" os 
\vspace{-3mm}
parameter
\be
\delta_{12}\,\propto \,-\ \Delta\left(\frac{N}{A_r}\right)_{\!12}\,.
\ee

\noindent
The $-$ sign corresponds to the fact that, for 	a new force effectively coupled to $B-L$, the test mass richer in neutrons (relatively to its mass) should undergo a stronger repulsive force from the neutrons in the Earth, leading to a smaller apparent ``free-fall'' acceleration.

\vspace{2mm}

This analysis further extends to situations involving two spin-0 doublets (allowing for an axial $U(1)_A$ generator $F_A$ to contribute to (\ref{f}))
and an additional singlet, breaking $U(1)_U$ so that the $U$ acquires a mass, here supposed to be extremely small \cite{npb81,plb80,plb86,plb89,newint}. The $U$ current then includes an axial part as well, strongly constrained by experimental results \cite{prd07}\!: 
indeed a light $U$ with non-vanishing axial couplings ($f_A$)  would interact with quarks and charged leptons very much as an axionlike pseudoscalar with effective pseudoscalar couplings $f_p=f_A \times 2m_{q,l}/m_U$, requiring $U(1)_U$ to be broken at  a sufficiently high scale through a large singlet v.e.v. \cite{npb81,plb80,plb86,plb89,newint}.  The vector part in the $U$ current, subject of our present interest, is still found to be a linear combination of the baryonic, leptonic and electromagnetic currents as given by (\ref{qu}), again involving $B-L\,$,
\,i.e. the number of neutrons $N$, in a grand-unified theory.

\section{Intensity of the new force, as compared to gravity}

The new interaction potential between two particles at distance $r$ reads
\be
\label{pot2}
V_U\,=\, \dis 
\frac{e^2}{4\pi\epsilon_\circ r} \  (\epsilon_Q Q +\epsilon_B B+\epsilon_L L)_1\, (\epsilon_Q Q +\epsilon_B B+\epsilon_L L)_2\,
\ e^{-r/\lambda_U} \,,
\ee
generalizing the corresponding expression for a pure dark photon.
The $U$ boson may be viewed as a generalized dark photon, including as well the 
gauge bosons of $B$ and/or  $L$, or $B-L$, and interpolating between these situations. It  takes into account the electroweak 
mixing effects with the photon and the $Z$, as illustrated by eqs.\,(\ref{33}-\ref{qu}).
\vspace{-.6mm}
For neutral objects with a small extension compared to $\lambda_U=\hbar/(m_Uc)$,
$
\,V_U=
\frac{e^2}{4\pi\epsilon_\circ r} \  (\epsilon_B B+\epsilon_L L)_1\, (\epsilon_B B+\epsilon_L L)_2\ e^{-r/\lambda_U},
$
reducing to a Coulomb-like potential
\vspace{.2mm}
for a sufficiently  long-ranged force.
For a coupling involving $Q$ and $B-L$, expressed 
as 
\be
[\,\epsilon_Q \,Q + \epsilon_{B-L}\,(B-L)\,] \ e\ 
\ee
 as expected from anomaly-cancellation, or grand-unification, $V_U$ reduces to \cite{plb89,plb86}
\be
V_U\,=\, \dis 
\frac{e^2}{4\pi\epsilon_\circ r} \  \,\epsilon_{B-L}^{\,2} \ N_1 N_2 \ \,e^{-r/\lambda_U}\, 
\ee
where $N=B-L$ denotes the number of neutrons.

\vspace{2mm}

Assuming $\lambda_U$  infinite, or at least somewhat larger than the Earth diameter,
the new potential between the Earth and a test body $i$ at a distance $r$ of its center
\vspace{-1mm}
 is in general
\be
V_U\,=\, \dis 
\frac{e^2}{4\pi\epsilon_\circ r}\  (\epsilon_B  B+\epsilon_L L)_\oplus\  (\epsilon_B B+\epsilon_L L)_i \,. \ee
Adding it to the attractive Newton potential
 $\,V_N=\,-\,G_N m_\oplus m_i/{r}\,$  amounts to the rescaling
\be
V_N\ \to \ V_N + V_U\,= \ (1+\delta_i)\ V_N\,.
\ee
The 
\vspace{-3mm}
ratio
\be
\delta_i \,=\,  \frac{V_U}{V_N\,}\,=\,-\  \frac{e^2}{4\pi\epsilon_\circ\,G_N m_p^2}\  (m_p/{\rm u})^2\
\frac{(\epsilon_B B+\epsilon_L L)_\oplus }{m_\oplus/{\rm u}}\ \frac{(\epsilon_B B+\epsilon_L L)_i}{m_i/{\rm u }}\,
\ee
may be expressed  in terms of the ratio of the electromagnetic to the gravitational forces between two protons, proportionally to
$\, {e^2}/{(4\pi\epsilon_\circ\,G_N m_p^2)}=  \alpha\, ({m_{\rm Planck}}/{m_p})^2\simeq 1.2356\ 10^{36}$.
\vspace{-.4mm}
Expressing masses in atomic mass units (the mass of a $^{12}$C  atom being 12\,u,  with
$m_p \simeq  1.007 \,276 $ u)
with $A_r$ denoting the relative atomic mass of an element, we get
\be
\ba{ccl}
\delta_i\,=\,\frac{\!\!\!\!\!\!\hbox{\phantom{g} new force}}{\hbox{gravity}}\,\simeq& 
 -\ 1.2536\ 10^{36}\ 
\left(\,\epsilon_B \,\hbox{\small$\dis\frac{B}{A_r}$}+\, \epsilon_L \, \hbox{\small$\dis\frac{L}{A_r}$}\,\right)_\oplus \  
\left(\,\epsilon_B \,\hbox{\small$\dis\frac{B}{A_r}$}+\,\epsilon_L \, \hbox{\small$\dis\frac{L}{A_r}$}\,\right)_i .
 \ea
 \ee
 
\vspace{2mm}
 
The $-$ expresses that the new force is in general repulsive,
while gravity is attractive. To ultimately obtain very small values of these parameters $\delta_i$, smaller than $10^{-12}$, in spite of the huge overall factor $10^{36}$, we  shall consider extremely small values of $\epsilon$ which are typically $\simle 10^{-24}$ in magnitude. 
Possible motivations for considering such small values of $\epsilon$ originating from extremely small values of $g"$ will be discussed in Section \ref{sec:hie} in the framework of supersymmetry \cite{ssm}, in connection with a very large hierarchy of mass or energy scales, typically between $\approx$ TeV scale and $\approx 10^{16}$ GeV scale or so. With $\epsilon^2 \simle 10^{-48}$ we can get  very small $|\delta_i|$'s $\,\simle 10^{-12}$, i.e. a new force smaller than gravity by 
about 12 orders of magnitude at least. The resulting apparent violations of the Equivalence Principle will then be typically 
 $\simle 10^{-15}$ for a new force coupled to $B$ (with $B/A_r$ very close to 1), up to a few $10^{-14}$ for a 
 coupling involving $L$, including most notably $B-L\,$. This will be made precised soon  in eq.\,(\ref{deltabis}), and (\ref{expdelta}) in the case of MICROSCOPE.

\section{Apparent violations of the Equivalence Principle}

Gravity seems to enjoy a remarkable universality property: bodies of different compositions fall at the same rate in an external gravitational field
\cite{micro,eot,adel1,adel2,adel3,adel4,lunar}.
Einstein interpreted this fact as an equivalence between gravitation and inertia, and used this Equivalence Principle as the starting point for the theory of General Relativity.
A new long-range force would lead to apparent violations of the Equivalence Principle, with changes in the observed accelerations of test bodies $i$ ``freely-falling'' towards the Earth, 
according 
\vspace{-3mm}
to
\be
g\ \to \ a_i \,= \, (1+\delta_i)\, g \,.
\ee

\noindent
This would imply
a non-vanishing value for the E\"otv\"os parameter measuring the relative difference in the observed accelerations of two test masses ``freely-falling'' towards the Earth,
\be
\delta_{12}\,= \,\frac{a_1-a_2}{(a_1+a_2)/2}\,\simeq \, \frac{a_1-a_2}{g}\,\simeq \ \delta_1-\delta_2\,,
\ee
\noindent 
i.e.
 \be
\label{deltaearth}
\delta_{12}\,\simeq\,
 -\ 1.2536\ 10^{36}\
\left(\,\epsilon_B \,\hbox{\small$\dis\frac{B}{A_r}$}+\, \epsilon_L \, \hbox{\small$\dis\frac{L}{A_r}$}\,\right)_\oplus\
\left[\ \epsilon_B \ \Delta\hbox{\small$\dis \left(\frac{B}{A_r}\right)$}_{\!12}\!+\, 
\epsilon_L \ \Delta\hbox{\small$\dis\left(\frac{L}{A_r}\right)$}_{\!12}\ \right]\,.
  \ee
These apparent violations of the Equivalence Principle could be rather large and even huge, unless  the new force is really very small compared to gravitation, and extremely small compared to the electromagnetic force.
For $|\epsilon|\simle 10^{-24}$,  they would be $\simle $ (1 to a few) $10^{-15}$ for a new force coupled to $B$, up to $\simeq $ 
\linebreak (1 to a few) $10^{-14}$ for a force coupled to $L$, or $B\!-\!L\,$ (cf. eqs.\,(\ref{deltabis},\,\ref{expdelta},\,\ref{expdelta2})), depending on $\Delta(B/A_r)$ and $\Delta(N/A_r)$.

\vspace{2mm}

For a neutral object $B$ and $L$  coincide with  the total numbers of nucleons ($B=A=Z+N$) and  protons ($L=Z$), with the mass, 
in atomic mass units, denoted $A_r$ as for a relative atomic mass.
 For an atom with $Z$ protons and $N_\alpha$ neutrons (with relative isotopic abundance $n_\alpha$, in amount fraction), the relative  atomic mass is $\,A_r=\Sigma \ n_\alpha A_{r\alpha}\,$, \,leading immediately to $L/A_r\,$. The average baryon number $B=$ $\Sigma \ n_\alpha (Z+N_\alpha)$ must be evaluated directly, to provide $B/A_r=\Sigma \ n_\alpha (Z+N_\alpha)\,/\,\Sigma \ n_\alpha A_{r\alpha}\,$, very close to 1.

\vspace{2mm}

To estimate $Z/A_r$ for the Earth, we consider its composition as 
32.1\% Fe ($Z/A_r\simeq .4656$), 30.1\% O ($Z/A_r\simeq .5000$), 15.1\% Si ($Z/A_r\simeq .4985$), 13.9\% Mg ($Z/A_r\simeq .4937$), 2.9\% S ($Z/A_r\simeq .4991$), 1.8\% Ni ($Z/A_r\simeq .4771$), 1.5\% Ca ($Z/A_r\simeq .4990$), 1.4\% Al ($Z/A_r\simeq .4818$), with 1.2\,\% of other elements. This leads in average to $Z/A_r\simeq .4870$. We find in a similar way $B/A_r\simeq 1.0008$, so that $N/A_r\simeq .5138$, and $Z/B\simeq .4866$, $N/B\simeq .5134$.
The E\" otv\" os parameter in (\ref{deltaearth}) may then be expressed 
as
  \be
  \label{delta}
 \framebox [12.3cm]{\rule[-.4cm]{0cm}{1cm} $ \dis
\delta_{12}\,\simeq \ -\ 1.2546\ 10^{36}\ \left(\epsilon_B + .4866\, \epsilon_L \right) \
\left[\,\epsilon_B \ \Delta \left(\frac{B}{A_r}\right)_{\!12}\!+\epsilon_L \ \Delta\left(\frac{Z}{A_r}\right)_{\!12}\,\right]\ ,
$}
\ee
reducing  to 
\vspace{-3mm}
  \be
  \label{deltabis}
  \left\{\ \ba{ccrc}\delta_{12\,B}\!&\simeq & -\ 1.2546\ 10^{36}& \epsilon_{B}^{2} \ \Delta \left( \hbox{$\dis\frac{B}{A_r}$}\right)_{\!12}\,,
\vspace{3mm}\\
\delta_{12\,L}\!&\simeq & -\ .6105\ 10^{36}& \epsilon_{L}^{2} \  \Delta \left( \hbox{$\dis\frac{L}{A_r}$}\right)_{\!12}\,,
\vspace{3mm}\\
\delta_{12\,B-L}\!&\simeq & -\ .6441\ 10^{36}& \epsilon_{B-L}^{\,2} \ \Delta \left( \hbox{$\dis\frac{N}{A_r}$}\right)_{\!12}\,,
\vspace{2mm}\\
\ea\right.
\ee
\vspace{-1mm}

\noindent
 for a coupling to $B,\ L$ or $B-L$\,.
 
 \vspace{2mm}
 As $\Delta(B/A_r)$ is usually close to 1 so that $\Delta(N/A_r)\simeq -\,\Delta(L/A_r)$, with the numbers of neutrons and protons within the Earth being approximately the same, the resulting expressions for $\delta_L$ (in terms of $\epsilon_L^2$) and $\delta_{B-L}$ (in terms of $\epsilon_{B-L}^{\,2}$) are usually approximately opposite as seen from (\ref{deltabis}), i.e.
\be
\label{deltadelta}
\frac{\delta_{12\,B-L}}{\epsilon_{B-L}^{\,2}}\,\approx \,-\ \frac{\delta_{12\,L}}{\epsilon_{L}^{2}}\ .
\ee 
This leads to approximately equal upper limits for $|\epsilon_L|$ and $|\epsilon_{B-L}|$, as long as no indication for a non-vanishing $\delta$ is found.
In general, 
 any upper limit on $|\delta|$ can be converted into an upper limit on $|\epsilon_B|,\ |\epsilon_L|$ or $ |\epsilon_{B-L}|$, for a force coupled to $B,\ L$ or $B-L\,$, according to
 \be
 \label{explim}
 \left\{\ 
 \ba{ccc}
 |\epsilon_B| <\,  .8928 \ 10^{-18} \  \hbox{$\dis \sqrt{\frac{\lim\,|\delta |}{|\Delta (B/A_r)|}}$}\, \ ,
   \vspace{2mm}\\
  |\epsilon_L| <\,   1.2798 \ 10^{-18} \ \hbox{$\dis \sqrt{\frac{\lim\,|\delta |}{|\Delta (L/A_r)|}}$}\, \ ,  \ \ \ \ 
   |\epsilon_{B-L}| <\, 1.2460 \ 10^{-18}\ \hbox{$\dis \sqrt{\frac{\lim\,|\delta |}{|\Delta (N/A_r)|}}$}\ .
 \ea \right.
 \ee
 The last two bounds have almost the same expressions, with  $\Delta (L/A_r)\simeq -\,\Delta (N/A_r)$ as indicated above, dealing with a new force acting effectively on either protons (for $L=Z$) or neutrons (for $B-L=N$). The bounds on $|\epsilon_L|$ and $|\epsilon_{B-L}|$ are thus almost the same as long as no indication for a non-vanishing $\delta$ is found, the bound on $|\epsilon_B|$ being typically several times larger, as $|\Delta(B/A_r)|$ is usually small.
 
 \section{Test-mass composition and E\" otv\"os parameter for MICROSCOPE}

We give in Table I the charge-to-mass ratios for the elements
composing the test masses of the MICROSCOPE experiment \cite{micro}. For $^{103}\rm Rh$ and $^{27}\rm Al$ 
we have a single isotope. 
The isotopic abundances of Pt ($A=190,\,192,\,194,\,195,\,196,\,198$) lead to  $A_r\simeq 195.084,\ B\simeq 195.120$.
For Ti ($A=46$ to 50) one gets  $A_r\simeq 47.867$ and
$B\simeq $ $47.9183$.
For Va (99,75\,\% $^{51}\rm Va$ and .25\,\% $^{50}\rm Va$), $A_r\simeq 50.9415$ and $B\simeq 50.9975$.
The resulting values for $Z/A_r,\ N/A_r$ and $B/A_r$ are shown in the last three lines of Table I.

  \vspace{2mm}
The MICROSCOPE test masses, however, are made of alloys: Platinum (90\,\%\,Pt\,--\,10\;\%\,Rh) and Titanium (TA6V, 90\,\% Ti\,--\,6\,\%\,Al\,--\,4\,\%\,V)
 alloys. 
 We average (with respect to the composition, in mass) the charge-to-mass ratios for pure bodies,
as shown in Table I. 
The resulting differences between test masses are given in Table II 
\,\footnote{The differences in Table II, evaluated with more digits than actually indicated, may thus differ by one unit in the last digit from 
those obtained from Table I.}.
Using alloys rather than pure Pt and Ti  does not modify these differences significantly. For example
$\Delta (N/A_r)\,\hbox{(Ti\,$-$\,Pt)}$  gets decreased by less than $5 \ \%$ in magnitude by  going to alloys,
from $\,-\,5.889\, \%\,$ to $-\,5.625\ \%$\,.

  \begin{table}
 \caption{  Charge-to-mass ratios $L/A_r$, $(B-L)/A_r$ and $B/A_r$, with $L=Z$ and $B=A=Z+N$ for neutral matter. 
 The values of $B$ are obtained by averaging on the isotopes with respect to their relative abundances (in amount fraction).
 The charge-to-mass ratios for the MICROSCOPE test masses, of Platinum  
 and TA6V Titanium alloys, are evaluated by averaging (with respect to their composition in mass) the values for pure bodies. 
 }
 \vspace{5mm}
$ 
 \ba{|c||c|c|c||c|c|c|c|}
 \hline
 &&&&&&& 
  \\ [-3mm] 
 &\ \ \ \ \ \hbox{Pt}\ \ \ \ \ &\ \ \ \ \  \hbox{Rh} \ \ \ \ \ &\ \ \ \hbox{Pt-Rh} \ \ \ &\ \ \ \ \ \hbox{Ti} \ \ \ \ \  &\ \ \ \ \  \hbox{Al} \ \ \ \ \ &
 \ \ \ \ \ \hbox{V} \ \ \ \ \  &\ \ \hbox{Ti-Al-V}\ \  
\\  & &&90 \%\,$-$\,10\%&&&& 90 \%\,$-$\,6\%\,$-$\,4\%
\\ [2mm] \hline \hline
 &&&&&&& \\ [-3mm]
L= Z& 78 & 45 &    & 22 & 13 & 23 & 
 \\  &&&&&&& \\ [-3mm] \hline 
&&&&&&& \\ [-3mm]
B &195.120&  103 & & \ 47.9183 & 27 & 50.9975 & 
 \\  &&&&&&& \\ [-3mm] \hline 
&&&&&&& \\ [-3mm]
A_r & 195.084 & 102.9055 &  & 47.867 &  26.9815 & 50.9415 & 
\\  &&&&&&& \\ [-3mm] \hline \hline 
&&&&&&& \\ [-3mm]
 L/A_r= Z/A_r & \ .39983 & \ .43729 & \  .40357 & \ .45961 & \ .48181 & \ .45150 & \ .46061
 \\  &&&&&&& \\ [-3mm] \hline 
 &&&&&&& \\ [-3mm]
 (B-L)/A_r= N/A_r&0 .60036 & \ .56362 & \ .59668 & \ .54146 & \ .51887 & \ .54960& \ .54043
 \\  &&&&&&& \\ [-3mm] \hline 
 &&&&&&& \\ [-3mm]
 B/A_r& 1.00018    
 & 1.00092 & 1.00026 & 1.00107 &1.00068  & 1.00110 & 1.00105
 \\  [-3mm]  &&&&&&& \\\hline
 \ea $
  \vspace{3mm}
 \end{table}

 \begin{table}
 \caption{  Differences in the charge-to-mass ratios $L/A_r$, $N/A_r$ and $B/A_r$ between the Ti and Pt alloys (and comparison with pure bodies).
$\Delta(L/A_r)$ and $\Delta(N/A_r)$, nearly opposite,  differ significantly from 0,  the Pt alloy being richer in neutrons by 
 $\simeq 5.6\ \%$ (with Pt richer than Ti by $\simeq 5.9\ \%$).
 }
 \vspace{3mm}
$ \ba{|c|c|c|}
\hline && \\  [-2mm]
&\  {\rm Ti}_{\hbox{\footnotesize alloy}} - {\rm Pt}_{\hbox{\footnotesize alloy}}\  & \ \ \ {\rm Ti} - {\rm Pt} \ \ \ 
 \\ [2mm]\hline
  && \\  [-3mm]
\ \Delta(L/A_r)\ &\ \  .05704 & \ \  .05978
\\  [2mm] \hline
 && \\  [-3mm]
\Delta (N/A_r) & \!\!\!- \ .05625 &   \!\!\! -\ .05889
\\ [2mm] \hline
 && \\  [-3mm]
\Delta(B/A_r)  & \ \ .00079  & \ \ .00089
\\  [2mm] \hline
\ea $
 \vspace{2mm}
\end{table}

\vspace{2mm}

We thus get for the E\" otv\"os parameter in the MICROSCOPE experiment, from eq.\,(\ref{delta}) and Table II,
  \be
  \label{delta2}
\delta(\rm Ti_{\hbox{\footnotesize alloy}},\,Pt_{\hbox{\footnotesize alloy}})\,\simeq \,-\,1.2546\ 10^{36}\, (\epsilon_B + .4866\  \epsilon_L ) 
\,(\,.00079\ \epsilon_B +.05704\ \epsilon_L )\ ,
\ee
not much different as  for pure bodies, as seen from Table II.
With the  $B/A_r$ ratios differing by less than $10^{-3}$, the E\" otv\" os parameter  is much more sensitive to $\epsilon_L$ than to $\epsilon_B$.
For couplings to $B,\ L$ or $B-L\,$, we find respectively:
\be
\label{expdelta}
\left\{\ \ba{ccrc}
\ \delta_B\ (\rm Ti_{\hbox{\footnotesize alloy}},\,Pt_{\hbox{\footnotesize alloy}})\!&\simeq &  -\ 1.00\ \ 10^{33}&\epsilon_B^2 \,,
\vspace{1mm}\\
\ \delta_L\ (\rm Ti_{\hbox{\footnotesize alloy}},\,Pt_{\hbox{\footnotesize alloy}})\!&\simeq &\!\!\! -\ 3.482 \ 10^{34}&\epsilon_L^2 \,,
\vspace{1mm}\\
\delta_{B-L}(\rm Ti_{\hbox{\footnotesize alloy}},\,Pt_{\hbox{\footnotesize alloy}}) &\simeq &\ \  \ \ 3.623 \ 10^{34}&\epsilon_{B-L}^2 \,,
\ea \right.
\ee
the last two expressions involving approximately opposite coefficients, as noted in (\ref{deltadelta}).
The positive value of $\delta_{B-L}$ indicates  that the Titanium alloy, less rich in neutrons, 
should undergo a smaller repulsive force from the neutrons in the Earth, leading to a larger $a$, and a positive $\delta_{B-L}$.
For $\delta_L$ the Ti alloy, richer in protons and electrons, should undergo a larger repulsive force from the electrons in the Earth, leading to a smaller $a$, and a  negative $\delta_L$.

\vspace{2mm}
In particular for a new force effectively coupled to $B\!-\!L$  we get from  (\ref{gbl}) 
\be
\label{gbl2}
\epsilon_{B-L}\,\simeq \,-\,\frac{5}{4} \,\frac{g"}{e}\,\simeq \,-\ 4.13 \ g"\,,
\ee
\vspace{-5mm} 

\noindent leading to
\be
\label{deltag}
\delta_{B-L}\,({\rm Ti_{\hbox{\footnotesize alloy}},\,Pt_{\hbox{\footnotesize alloy}}})\  \simeq \ 3.623 \ 10^{34}\ \epsilon_{B-L}^2
 \simeq\ .617\ 10^{36}\,g"^2 \,,
\ee
so that any upper limit on $\delta$ can directly be converted into an upper limit on $|\,g"|$.

\section{Constraints from Microscope}

We just got the first results of the MICROSCOPE experiment \cite{micro}, which implies  
\be
\label{micro}
\delta\,(\rm Ti_{\hbox{\footnotesize alloy}},Pt_{\hbox{\footnotesize alloy}})\,=\,[\,-.1\pm.9 \ (stat)\pm .9\  (syst)\,]\ 10^{-14}\,,
\ee
i.e. $(\,-.1\pm 1.3)\,\ 10^{-14}$  and $(\,-.1\pm 2\,)\ 10^{-14}$ at the 1$\sigma$ and 2$\sigma$ levels, respectively, when systematic and statistical errors are added in quadrature.
Considering that  $\,-\,1.4\  10^{-14}< \delta <1.2\  10^{-14}$\,
at 1$\sigma$,
 \,or $-\,2.1\  10^{-14}< \delta  <1.9\  10^{-14}$ 
at 2$\sigma$, we get from (\ref{expdelta},\ref{micro})
 the constraints
 \be
 \label{limite1}
 \left\{\,
 \ba{ccrl}
\hbox{for a force coupled to}& B\,: & |\epsilon_B|\, & <\ 3.7\ 10^{-24}\,,
\vspace{1mm}\\
\hbox{for a force coupled to} & L\,: & |\epsilon_L|\, & <\ .63\ 10^{-24}\,,  \ \  (1\,\sigma)
\vspace{1mm}\\
\hbox{for a force coupled to} & B-L\,:& |\epsilon_{B-L}|\!& <\ .57\ 10^{-24}\,, 
\ea \right.
\ \ \ \hbox{or}\ \ \
 \left\{\,
\ba{rl}
|\epsilon_B|\, & <\ 4.6\ 10^{-24}\,,
\vspace{1mm}\\
| \epsilon_L|\, & <\ .78\ 10^{-24}\,,  \ \  (2\,\sigma)
\vspace{1mm}\\
|\epsilon_{B-L}|\!& <\ .73\ 10^{-24}\,.
\ea\right.
\ee
 This may be remembered as
\be
\label{limite}
  \framebox [9.8cm]{\rule[-.25cm]{0cm}{.7cm} $ \dis
|\,\epsilon_B\,| \, <\,5\ 10^{-24}\,, \ \ \  |\,\epsilon_L\,|  \ \hbox{or}\  
|\,\epsilon_{B-L}\,| \,< \, .8\ 10^{-24}\,\ \ \ \ (2\sigma)\ . 
$}
\ee

\vspace{2mm}
In particular for a new force effectively coupled to $B-L$, for which we also have
\be
|\,\epsilon_Q\,| \,\simeq \, .61\ |\,\epsilon_{B-L}\,| \,<\ .45 \ 10^{-24}\,,
\ee
we get from  (\ref{gbl},\ref{gbl2})  $\,g"\simeq -\,\frac{4}{5} \,\epsilon_{B-L}\, e \,\simeq  \,-\,.242\ \epsilon_{B-L}  $, \,leading to
\be
  \framebox [3.8cm]{\rule[-.2cm]{0cm}{.6cm} $ \dis
|\,g"|\,<\, .18\ 10^{-24}\,,
$}
\ee
as obtained directly from (\ref{deltag}).
We recall that the sign of the extra-$U(1)$ gauge coupling constant $g"$ can be conventionally defined at will
as being $\,+$\,, or $\,-\,$, \,thanks to the possibility of redefining the extra-$U(1)$ gauge field $C$ through a change of sign.
These limits may be improved in the near future, possibly by a factor up to about 3 or so from a gain of sensitivity of the experiment by about an order of magnitude, if no positive indication in favor of a non-vanishing $\delta$ is found.

\vspace{2mm}

With the MICROSCOPE satellite orbiting at a $z\simeq 710$ km altitude, the limits obtained are valid for a range $\lambda_U$ 
in (\ref{range}) somewhat larger than the diameter of the Earth.  They become less constraining for smaller values of $\lambda_U$. For a range significantly smaller than 710 km the experiment gets nearly insensitive to such extremely small forces as $\delta $ now includes a very small global damping factor $e^{-\,z/\lambda_U}$, with the resulting upper limits on $|\epsilon|$ and $|\,g"|$ increasing very rapidly, much like $e^{z/(2\lambda_U)}$, for smaller values of $\lambda_U$, as compared to $z$.

\section{Comparison with earlier experiments}

These limits may be compared with those that may be derived from \cite{adel3,adel4}
\be
\delta \,\hbox{(Be,Ti)}= (3\pm 18)\ 10^{-14}\ \ \hbox{and} \ \ 
\delta \,\hbox{(Be,Al)}= (-7\pm 13)\ 10^{-14}\ \  (\hbox{at } 1\sigma) \ .
\ee
\vspace{-3mm}
With 
\be
\hbox{(Be,\,Ti):}\   
\left\{\, \ba{ccr}
\Delta (B/A_r)\!&\simeq&\! -\, .242 \,\%\,,
\vspace{.5mm}\\
\Delta (L/A_r)\!&\simeq&\! -\,1.577\,\%\,,
\vspace{.5mm}\\
\Delta (N/A_r)\!&\simeq&\!  1.333\,\%\,,
\ea\right.
\ \ \ 
\hbox{(Be,\,Al):}\   
\left\{\, \ba{ccr}
\Delta (B/A_r)\!&\simeq&\! -\, .203 \,\%\,,
\vspace{.5mm}\\
\Delta (L/A_r)\!&\simeq&\! -\,3.797\,\%\,,
\vspace{.5mm}\\
\Delta (N/A_r)\!&\simeq&\!  3.593\,\%\,,
\ea\right.
\ee   
we get from (\ref{deltabis}) 
\be
\label{expdelta2}
\left\{\, \ba{ccrc}
\ \delta_B\ (\rm Be,Ti)\!&\simeq &  3.04\  10^{33}&\epsilon_B^2 \,,
\vspace{.5mm}\\
\ \delta_L\ (\rm Be,Ti)\!&\simeq &\ .963 \ 10^{34}&\epsilon_L^2 \,,
\vspace{.5mm}\\
\delta_{B-L}(\rm Be,Ti) &\simeq &- \ .859\ 10^{34}&\epsilon_{B-L}^2 \,,
\ea \right.
\ \ \
\left\{\, \ba{ccrc}
\ \delta_B\ (\rm Be,Al)\!&\simeq &  \ 2.55\  10^{33}&\epsilon_B^2 \,,
\vspace{.5mm}\\
\ \delta_L\ (\rm Be,Al)\!&\simeq &2.32\   10^{34}&\epsilon_L^2 \,,
\vspace{.5mm}\\
\delta_{B-L}(\rm Be,Al) &\simeq & -\ 2.31 \ 10^{34}&\epsilon_{B-L}^2 \,.
\ea \right.
\ee
 The $\delta_{B-L}$'s are negative as the Be test masses are comparatively richer in neutrons than Ti or Al ones, with 5 neutrons for 4 protons. In the  opposite way the $\delta_L$'s are positive as the Be masses are comparatively less rich in protons, and thus electrons. 
 A positive $\delta $ is taken to be less than  $21\ 10^{-14}$ and $13\ 10^{-14}$  for Be-Ti and Be-Al, respectively. For a negative delta, $|\delta|$  is taken to be less than
 $18\ 10^{-14}$ and $20\ 10^{-14}$, respectively. One then gets the 
(1$\sigma$) upper limits  
\be
\label{limitebe}
\hbox{(Be,\,Ti):}\   \left\{\, \ba{ccr}
|\epsilon_B| \!&<&\!8.3\ 10^{-24}\,,
\vspace{.5mm}\\
|\epsilon_{L}| \!&<&\!4.7\ 10^{-24}\,,
\vspace{.5mm}\\
|\epsilon_{B-L}| \!&<&\!4.6\ 10^{-24}\,,
\ea \right.
\ \ \ \hbox{(Be,\,Al):}\  
\left\{\,\ba{ccr}
|\epsilon_B| \!&<&\!7.2\ 10^{-24}\,,
\vspace{.5mm}\\
|\epsilon_{L}| \!&<&\!2.4\ 10^{-24}\,,
\vspace{.5mm}\\
|\epsilon_{B-L}| \!&<&\!3\ \ \, 10^{-24}\,,
\ea \right.\ \ \ (1\sigma)\ 
\ee
in agreement with (\ref{explim}), and with the conservative estimate $|\,\epsilon_{B-L}\,| \simle10^{-23}$
given in \cite{epjc}.

\vspace{2mm}

Generally speaking, improving by one order of magnitude the sensitivity of the Equivalence Principle tests
 allows for the upper bounds on $|\epsilon|$, and $|\,g"|$, to be improved by a factor $\simeq 3$\,. 
 The first results of the MICROSCOPE experiment already allow for  such an improvement  by a factor $\simeq $ 4 to 5 for the upper bounds on $|\epsilon_L|$ or 
 $|\epsilon_{B-L}]$, as seen from  the comparison between (\ref{limite1}) and (\ref{limitebe}). The improvement  factor is smaller  for $|\epsilon_B|$,  due to a larger $|\Delta (B/A_r)|$, slightly above 2 $10^{-3}$, for the Be-Ti and Be-Al experiments, as compared with less than $10^{-3}$ for Ti-Pt in MICROSCOPE.

  \section{\boldmath Very small couplings \,from a large  \vspace{1.8mm }hierarchy, 
 \hbox{as associated with \,inflation} \,and \,supersymmetry breaking}
 \label{sec:hie}
 
 \subsection{\boldmath An extremely small  $g"$, \,with a very large energy scale from a $\,\xi" D"$ term}
 
 \vspace{-1mm}
 
What are the implications of these results, and why should we care about such extremely small gauge couplings, typically $\simle 10^{-24}$\,? 
\,As we shall see such small values may be related with a correspondingly large hierarchy in mass or energy scales, by a factor  $\simge 10^{12}$, involving a very large scale of the order of 
\linebreak $\sim \,10^{16}$ GeV.

\vspace{2mm}

A crucial role is played, within supersymmetric theories of particles,
by the $\xi D$ term \cite{fi} present in the Lagrangian density for an invariant $U(1)$ gauge subgroup, such as the $U(1)_F$  considered here.
 \,Indeed we considered long ago  the limit of a vanishing gauge coupling, \hbox{$g"\to 0\,$}, for an extra $U(1)$ gauge group beyond $SU(3)\times SU(2)\times U(1)_Y$ \cite{ssm}. This was done in connection with the question of spontaneous supersymmetry breaking, with a very large $\xi"D"$ term in the Lagrangian density associated with the extra $U(1)$ factor in the gauge group.  Boson mass$^2$ terms of moderate size, proportional to $\xi"g"$,  may then be obtained by combining the very large coefficient $\xi"$ having the dimension of a mass$^2$ with the very small $U(1)$ gauge coupling $g"$.
These terms are obtained from the contribution to the scalar potential
 \be 
 \label{v}
 V"\,=\,\hbox{$\dis\frac{D"^2}{2}$}\,=\,\frac{1}{2}\ |\,\xi"\,+\,\hbox{$\dis\frac{g"}{2}$}\ 
 \hbox{\footnotesize $\dis\sum_i$}\ F_i\ \varphi_i^\dagger\varphi_i\,|^2\,=\ \hbox{$\dis\frac{\xi"^2}{2}$}\,+\ \hbox{\footnotesize $\dis\sum_i$}\ 
 \,\hbox{$\dis \frac{\xi"g"}{2}$} \,F_i\ \varphi_i^\dagger\varphi_i\,+\  ...\ \,.
 \ee
 $\varphi_i$ denote the spin-0 components of left-handed chiral superfields $\Phi_i$, with 
couplings $(g"/2)\,F_i$ to the extra $U(1)$ gauge field $C^\mu$.
This provides the soft supersymmetry-breaking spin-0 boson mass$^2$ terms
 \be
 \label{mui}
 \mu_i^2\,=\,\,\hbox{$\dis \frac{\xi"g"}{2}$} \ F_i\ .
 \ee
 Taking the limit $\xi"\to\infty$, $g"\to 0$\, with $\,\xi"g"/2=\mu_\circ^2$ fixed, corresponding to supersymmetry getting spontaneously broken ``at a very high scale" \cite{ssm}, still generates the above dimen\-sion-2 soft supersym\-metry-breaking terms (\ref{mui}).  Then $\xi"$ appears to be 
 proportional to $ 1/g"$ and very large, or conversely, 
 $g"\propto 1/\xi"$, and very small.
 
 \vspace{2mm}

The auxiliary field $D"$ contributes to an initial vacuum energy density
 through the term 
\be
\label{V0}
\framebox [5.2cm]{\rule[-.3 cm]{0cm}{.9cm} $ \dis
V"_{\!\!\circ}\ = \ \frac{\xi"^2}{2}\,\propto\ \frac{1}{g"^2}\ \ \ \hbox{(huge)}\,,
$}
\ee
originating from expression  (\ref{v}) of the potential. This term may be responsible for a very rapid inflation of the early Universe, typi\-cally 
 \vspace{-2mm}
 with
 \be
 \xi"\ \approx \ \Lambda_{\rm inflation}^2\,.
 \ee
More precisely with $\,{\cal L} = {D"^2}/{2} + \xi"\,D" \!+ ... \,$,
\,$\xi"$  enters in the equation of motion for the auxiliary field, $\,D"\!=-\,\xi"+ ... $\,, so that $D"$, \,equal to $ -\,\xi"$ when all spin-0 fields vanish, may acquire a different v.e.v.  $\,<\!D"\!>\ = -\,\xi"_{\!\rm eff}\,$, with in general 
$\,|\!<\!D"\!>\!|\,$ decreased from $|\,\xi"\,|$ to $|\xi"_{\!\rm eff}|$.
$\ \xi"$ should then ultimately be replaced by $\xi"_{\!\rm eff}$,  which, when non-zero so as to contribute to supersymmetry breaking, is expected to be proportional to $\xi"$ and in many cases of the same order, but smaller in magnitude. Still we also have to pay attention that the $U$ boson should not acquire a too large mass in this process. This may be achieved, in particular, if  the shift in $<\!D"\!>\,$ from $-\,\xi"$ to $-\,\xi"_{\!\rm eff}$ originates from the very large v.e.v. $\,<\!\varphi^\dagger\varphi\!>\,$  of a bilinear form involving strongly-coupled (for a new interaction with a very large energy scale) spin-0 fields $\varphi$, significantly coupled to $C^\mu$ in a hidden sector, 
\,still with $<\!\varphi\!>\ =0\,$ so that this effect does not break the extra-$U(1)$ symmetry.

 \vspace{2mm}
The new v.e.v. $<\!D"\!\!>\ =-\,\,\xi"_{\!\rm eff}\,$  replacing the original $-\,\xi"$ contributes to the generation, for the other spin-0 fields of interest to us, of super\-sym\-metry-brea\-king  mass$^2$ terms expressed as 
\be
\mu_i^2\,=\,\,\hbox{$\dis\frac{\xi"_{\!\rm eff} \,g"}{2}$}\ F_i\,=\,\mu_{\circ\,\rm eff}^{\,2} \,F_i\,,
\ee
obtained from the expansion of  (\ref{v}).
We can thus express the very small extra-$U(1)$ gauge coupling as
 \be
  \label{g"}
 g"=\ 
 \frac{2\,\mu_{\circ\,\rm eff}^{\,2}}{\xi"_{\!\rm eff}}\ .
 \ee
 The parameter $\mu_{\circ\,\rm eff}^{\,2}= \xi"_{\!\rm eff} \,g"/2\,$ is normally expected  to be of the order of the boson-fermion mass$^2$-splittings within the multiplets of super\-symmmetry, now usually believed $\,\approx  (1\!-\!10$ TeV)$^2$,
 \vspace{-2mm}  leading to
 \be
 \label{g"2}
 |\,g"|\,\approx\, \frac{m_{\rm sparticle}^{\,2}}{|\,\xi"_{\!\rm eff}|}\,\approx\, \frac{(1\!-\!10 \ \hbox{TeV})^{2}}{|\,\xi"_{\!\rm eff}|}\ .
 \ee
 
 \vspace{2mm}

 In practice considering $g" \to 0$ and $\,\xi"\to\infty$ means considering very large values of $\,\xi"$, and often of $\xi"_{\!\rm eff}$ as well so that it contributes effectively to the supersymmetry breaking in the final state, $\,\approx \Lambda^2$. 
 This $\Lambda$  may be one
 of the very large scales in the theory, such as the grand-unification scale or  inflation scale (taken 
$ \approx $ $10^{16}$ GeV), a compactification scale
\vspace{-.5mm}
 (possibly related to the grand-unification scale by \linebreak 
 $m_X\approx \pi\hbar/\,Lc$~\cite{f87}), a very large supersymmetry-breaking scale $\Lambda_{\rm ss}\approx \sqrt{|\xi"_{\!\rm eff}}|\approx \sqrt{\phantom {l}\!m_{3/2} \,m_{\rm Planck}}\,$ 
 \vspace{-.5mm}
 associated with a very heavy gravitino, or a string scale $\,\approx 10^{17}\!-\!10^{18}$ GeV. \,This then leads to
\be
\label{g"lambda}
|\,g"|\,\approx\, \left(\frac{m_{\rm sparticle}}{\Lambda}\right)^{\!2}\,\approx\, \left(\frac{1\!-\!10 \ \hbox{TeV}}{\Lambda}\right)^{\!2}\,\approx\ 10^{-26}\!-\!10^{-24}\ ,
\ee

\vspace{-2mm}

\noindent
for $\Lambda\approx 10^{16}$ GeV.
\vspace{2mm}

With for example 
$
\,\delta_{B-L}\,({\rm Ti_{\hbox{\footnotesize alloy}},\,Pt_{\hbox{\footnotesize alloy}}})  \simeq \,3.62 \ 10^{34}\ \epsilon_{B-L}^2
 \simeq\, .62\ 10^{36}\,g"^2$
as in (\ref{deltag}),
we get from (\ref{expdelta}) the possible order of magnitude estimate for the E\" otv\" os parameter,
\be
\label{deltag"}
\framebox [13.2cm]{\rule[-.45cm]{0cm}{1.1cm} $ \dis
\delta\ \simeq\ .6\ 10^{36}\,g"^2
\ \approx \, 10^{36} \,  \left(\frac{m_{\rm sparticle}}{\Lambda}\right)^{\!4}\,\approx\, 10^{36} \,  \left(\frac{1\!-\! 10\ \hbox{TeV}}{\Lambda}\right)^{\!4}\,\approx\ 10^{-16}\!-\! 10^{-12}\,,
$}
\ee
for $\Lambda \approx 10^{16}$ GeV. 

\vspace{2mm}

This estimate decreases by 4 orders of magnitude, down to  $\approx 10^{-20}-10^{-16}$ for 
$\Lambda \approx 10^{17}$ GeV. 
A constraint  of less than about $2 \ 10^{-14}$ on $|\delta|$, as obtained at $2\sigma$ from the MICROSCOPE results, thus already provides useful information on an extremely weak long-range interaction associated with a very large energy 
scale. To fix the ideas, with sparticle masses of the order of 3 TeV$/c^2$ and an inflation scale around $10^{16}$ GeV, we might expect
in this framework apparent violations of the Equivalence Principle  with an E\"otv\"os parameter $\delta$ roughly of the order of $10^{-14}$, not much below the present MICROSCOPE limit
 \cite{micro}.

\subsection{A huge vacuum energy density \boldmath $\,\propto 1/g"^2$,  \vspace{1.8mm} \,the inflation of the Universe, 
\hbox{and supersymmetry breaking with a very heavy gravitino}}
  \vspace{-1mm}
  
  The term ${\xi"^2}/{2}\,$ in (\ref{V0}),  from expression (\ref{v}) of the potential,
 appears as the very large contribution of the $\xi"D"$ term in $\cal L$ to the vacuum energy density, evaluated when all spin-0 physical fields vanish. It  may be responsible, by itself  or jointly  with a linear superpotential term
 $\sigma S$,
    \vspace{-.6mm}
  for the very rapid inflation of the early Universe,
 with a huge contribution to its initial vacuum energy density 
  $V_\circ= \frac{\xi"^2}{2}+|\sigma ^2|$. This connection with inflation, from $D$ and $F$ terms in the potential, 
  associated with a very 
  small extra-$U(1)$ coupling $g"$ and  a very light $U$ boson that could lead to a  new long range force, was already pointed out long ago \cite{f84}. 
  
 \vspace{2mm} 
 
  The contribution $D"^2/2$ to the vacuum energy density can decrease quickly  from the initial $\,\xi"^2/2$ to a lower $\xi"_{\!\rm eff}^{2}/2$ as the result of a dynamical modification of the $\xi"D"$ term, possibly induced by the translation of a bilinear form in spin-0 fields, $<\!\varphi^\dagger\varphi\!>$, \,strongly coupled in a hidden sector.
 The v.e.v. of auxiliary $F$ components of chiral superfields, when present, may also be modified  (or simply generated) through the change
  $|\sigma| \to |\sigma_{\rm eff}|$. We ignore here possible contributions from weak-hypercharge and weak-isospin auxiliary $D$ fields, which play no significant role in this discussion.
  The Universe gets ultimately almost flat, but for an extremely small residual cosmological constant,
thanks to the supergravity contribution $- \,3  \,m_{3/2}^2\,m_{\rm Planck}^2/8\pi\,$ 
cancelling out in the final vacuum state,
    \vspace{-.4mm}
 almost completely, the 
$\,d^2/2=\xi"_{\!\rm eff}^2/2\ +|\sigma_{\rm eff}|^2\,$ energy-density contribution from $D$ and $F$ terms \cite{dz,grav,f84}.
\vspace{-1mm}
\,With
\be
(\xi",\ \sigma)\ \, \to\, \ (\xi"_{\!\rm eff},\ \sigma_{\rm eff})\,,
\ee
during a rapid phase of expansion of the Universe,
this change in the vacuum energy density $\rho_{\rm vac}$ may be expressed as  
\,\footnote{The first extensions of the supersymmetric standard model \cite{ssm}, with spontaneously broken supersymmetry, were of this type,  with $\xi"_{\!\rm eff}\neq 0$ and $\sigma_{\rm eff}\neq 0$ both generated from the extra-$U(1)$ $\,\xi"D"$ term in the Lagrangian density. This  term is then alone responsible for the inflation of the Universe, in the absence of a linear 
$\sigma S$ superpotential term forbidden by the extra-$U(1)$ symmetry acting according to
$S\to e^{-2i\alpha} S,\ H_1\to e^{i\alpha} H_1$, $H_2\to e^{i\alpha} H_2$,\, leaving the trilinear $\lambda\,H_2H_1S$ superpotential term invariant.
}
\be
\label{vac}
\rho_{\rm vac}^i= \,\frac{\xi"^2}{2} +|\sigma|^2- \,3  \,m_{3/2}^{i\ 2}\,{m_{\rm Planck}^2}/{8\pi} \ \ \ \ \searrow \ \ \ \ 
\rho_{\rm vac}^f= 
\,\frac{\xi"_{\!\rm eff}^2}{2} +|\sigma_{\rm eff}|^2- \,3  \,m_{3/2}^2\,{m_{\rm Planck}^2}/{8\pi}\,\simeq \,0\,.
\ee
The $\simeq $ 0 for $ \rho_{\rm vac}^f$ may be replaced by $\Lambda_{\rm cc} c^4/(8\pi G_N)$ in which $\Lambda_{\rm cc}$ is the value, extremely  small, of the cosmological constant taken as responsible for the present acceleration of the Universe. 
\vspace{2mm}

The inflation scale is then defined as
\vspace{-1.5mm} 
\be
\label{infl}
\Lambda_{\rm inflation}^{\,4}\,=\,\rho_{\rm vac}^i\!-\rho_{\rm vac}^f\,\simeq\,\frac{\xi"^2-\,\xi"_{\!\rm eff}^2}{2}+ |\sigma|^2-|\sigma_{\rm eff}|^2\,,
\ee
assuming for simplicity that the gravitino mass does not get significantly modified during this early expansion phase of the Universe.
Assuming also  that $\,\xi"_{\!\rm eff}^2$ is of the same order as  
$\Lambda_{\rm inflation}^4$ 
and combining 
 (\ref{g"2}) and (\ref{infl}) 
 we get as in (\ref{g"lambda}) 
 \vspace{-3mm}
\be
|\,g"|\,\approx\, \left(\frac{m_{\rm sparticle}}{\Lambda_{\rm inflation}}\right)^{\!2}\,,
\ee
leading to the estimate (\ref{deltag"}) for the E\"otv\"os parameter $\delta$.

\vspace{2mm}

Let us now consider the mass of the gravitino, fixed in flat (or almost flat) spacetime by the product of the gravitational coupling
$\kappa= \sqrt {8\pi\,G_N}$ with the supersymmetry-breaking scale parameter, defined by
$d^2/2=F_{\rm ss}^2=
\sum <\!D\!>\!\!^2/2\  + $
$\sum |\!\!<\!F\!>\!\!|^2= |\,\xi"_{\!\rm eff}|^2/2 +\,|\sigma_{\rm eff}|^2\,$, leading to
\be
m_{3/2}\,=\,\frac{\kappa\,d}{\sqrt 6}\,=\,\frac{\kappa \,F_{\rm ss}}{\sqrt 3}\,.
\ee
This relation also ensures that the $\pm 1/2$ polarisation states of a very light spin-3/2 gravitino have enhanced gravitational interactions, and behave very much as a massless spin-1/2 goldstino with non-negligible gauge couplings $\approx g"$ \cite{grav} (but this is not the situation in which we are interested here). It implies that the present energy density of the vacuum $\rho_{\rm vac}^f$ in (\ref{vac}) cancels out, almost exactly but for an extremely small contribution associated with a non-vanishing cosmological constant $\Lambda_{\rm cc}\,$.

\vspace{2mm}
We assume $\xi"_{\!\rm eff} \neq 0$ so that it contributes effectively to spontaneous supersymmetry breaking, and express it relatively to $d$  as in \cite{grav}, according to
$|\,\xi"_{\!\rm eff}|= $ $d\,\cos\theta"$.
It leads, within supergravity (including if needed a possible contribution of the $R$-symmetry current to the 
$U(1)_F$ current associated with (\ref{f})),
 to a gravitino  
 \vspace{-2mm}
 mass 
\be
m_{3/2}\,=
\,\frac{\kappa\,d}{\sqrt 6}= \,\sqrt\frac{4\pi}{3}\ \frac{|\,\xi"_{\!\rm eff}|}{m_{\rm Planck }\cos\theta"}\,\approx\,\frac{|\,\xi"_{\!\rm eff}|}{m_{\rm Planck}}\ \,,
\ee
with $\kappa
= \sqrt{8\pi}/m_{\rm Planck}\simeq 4.1 \ 10^{-19}\ \hbox{GeV}^{-1}$. 
\vspace{-.5mm}
\,With $\, |\,\xi"_{\!\rm eff}| \approx m_{\rm sparticle}^{\,2}/{|\,g"|}$ from (\ref{g"2}) we obtain a relation 
between the
gravitino mass, the gravitational coupling $\kappa$,
the boson-fermion mass$^2$-splittings and the extra-$U(1)$ coupling $g"$, expressed as in \cite{grav}
 \vspace{-1mm}
 according to
\be
m_{3/2}\,\approx\,\kappa\ \frac{m_{\rm sparticle}^{\,2}}{|\,g"| \cos\theta"}\ .
\ee
A very heavy gravitino is then naturally associated with a very small gauge coupling (while, conversely,  a very light one would have a goldstino-like behaviour with non-negligible gauge couplings).

\vspace{2mm}

A very large $\xi"_{\!\rm eff}$, associated with a very small  $g"$ as considered in  \cite{ssm,grav}, corresponds to a very large supersymmetry-breaking scale
\be
\Lambda_{\rm ss}=\sqrt {F_{ss}}=\frac{\sqrt d}{2^{1/4}}\,=\,
\left(\frac{3}{8\pi}\right)^{\!\!1/4}\! \sqrt{{\phantom{\hbox{\small  i}}}m_{3/2}\ m_{\rm Planck}}
\ =\,\frac{|\,\xi"_{\!\rm eff}|^{1/2}}{2^{1/4}\, \cos^{1/2}\theta"}\ \,,
\ee
with a heavy gravitino,  leading to  
\be
|\,g"|\,\approx\  \frac{m_{\rm sparticle}^{\,2}}{m_{3/2}\ m_{\rm Planck}}\approx  \left(\frac{m_{\rm sparticle}}{\Lambda_{\rm ss}}\right)^{\!2}  \ .
\ee
And, using (\ref{deltag}) to estimate $\delta$ from $g"$, to
\be
\framebox [9.2cm]{\rule[-.4cm]{0cm}{1.05cm} $ \dis
\delta
\, \approx \, 10^{36}  \left(\frac{m_{\rm sparticle}^{\,2}}{m_{3/2}\ m_{\rm Planck}}\right)^{\!2}
\, \approx \, 10^{36}  \left(\frac{m_{\rm sparticle}}{\Lambda_{\rm ss}}\right)^{\!4}\,.
$}
\ee
Taking $\,m_{3/2}\approx 10^{13}$ GeV$/c^2$, i.e. supersymmetry broken ``at the very large scale" $\Lambda_{\rm ss}\approx \sqrt{{\phantom{\hbox{\small  i}}}m_{3/2}\,m_{\rm Planck}}$ $\approx 10^{16}$ GeV, with $m_{\rm sparticle} \approx$ 1 to 10 TeV/$c^2$, leads again 
 to 
$
\,g" \approx10^{-26}\!-\!  10^{-24}$ as in (\ref{g"lambda}),
and to an E\" otv\" os 
\vspace{-2mm}
parameter
\be
\delta\,\approx \,10^{-16}\!-\!10^{-12}\,,
\ee
decreasing to $\delta \approx  10^{-18} \!-\! 10^{-14}$ for $m_{3/2}\approx 10^{14}$ GeV/$c^2$. 

\vspace{3mm}
\bc

* \hspace{8mm}  *

\vspace{4mm}
*
\ec

\vspace{2mm}
    
The first results of MICROSCOPE, constraining the Ti-Pt E\" otv\" os parameter $|\delta|$ to be less than about $2\ 10^{-14}$ at 2$\sigma$, confirm that a new long-range force added to gravity must be extremely weak, typically with a gauge coupling $g"$ smaller than $10^{-24}$, providing increased constraints on its possible magnitude. Testing the Equivalence Principle to a very high degree of precision may be  the most powerful way to look for extremely feeble long-range forces, complementing the search for very heavy particles and short-range interactions performed at LHC.
 
\vspace{2mm} 
 Should such a force be observed, its characteristics may allow to find out if it is due to a spin-1 $U$ boson mediator, effectively coupled to $B$ and $L$, and help shedding light  on a possible unification of weak, electromagnetic and strong interactions.
 The extremely small couplings tested, down to less than $10^{-24}$, may be related to a very large hierarchy of mass scales by a factor $\approx 10^{12}$ at least, involving the ratio of a moderate scale $\approx$ a few TeV's as tested in LHC to a very large one typically  $\approx 10^{16}$ GeV, as may be associated with grand-unification or compactification, inflation, or supersymmetry-breaking with a very heavy gravitino.

\bibliography{References}

\end{document}